\documentclass[runningheads,a4paper]{llncs}

\usepackage{amssymb}
\usepackage{graphicx}
\usepackage{url}
\urldef{\mailsa}\path|{x.d.arsiwalla@gmail.com}| 
\newcommand{\keywords}[1]{\par\addvspace\baselineskip
\noindent\keywordname\enspace\ignorespaces#1}

\begin{document}

\mainmatter  

\title{Measuring the Complexity of Consciousness } 


\author{Xerxes D. Arsiwalla\inst{1,2,3}%
\and Paul Verschure\inst{1,2,3,4}}
%

\institute{Institute for Bioengineering of Catalonia, Barcelona, Spain. \\ 
\and
Barcelona Institute for Science and Technology, Barcelona, Spain.  \\ 
\and 
Universitat Pompeu Fabra,  Barcelona, Spain. \\ 
\and
Instituci\'{o} Catalana de Recerca i Estudis Avan\c{c}ats (ICREA), \\ 
Barcelona, Spain.\\
\mailsa\\ 
}


\pagestyle{empty}
\maketitle

\begin{abstract}
The quest for a scientific description of consciousness  has given rise to new theoretical and empirical paradigms for the investigation of phenomenological contents   as well as clinical disorders of consciousness.  An outstanding challenge in the field is to develop measures that uniquely quantify global brain states tied to consciousness.   In particular,  information-theoretic complexity measures such as integrated information have recently been proposed as measures of conscious awareness. This  suggests a new framework to quantitatively classify states of consciousness.  However, it has proven increasingly difficult to apply these complexity measures to realistic brain networks. In part, this is due to high computational costs incurred when implementing these measures on realistically large network dimensions.  Nonetheless,  complexity  measures for quantifying states of consciousness are important for assisting clinical diagnosis and therapy.  This article is meant to serve as a lookup table of measures of consciousness, with particular emphasis on clinical applicability of these measures.  We consider both, principle-based complexity measures as well as empirical measures tested on patients.  We address challenges facing these measures with regard to realistic brain networks, and  where necessary,  suggest possible resolutions.   

\end{abstract}

\section{Introduction }  
In patients with disorders of consciousness,  such as coma, locked-in syndrome or   vegetative state,  levels of consciousness are assessed in the clinic through a battery of behavioral tests and neurophysiological recordings.  In particular, these  methods are used to assess  levels of  wakefulness (arousal) and awareness in patients \cite{laureys2004}, \cite{laureys2005}.  Such assessments have led to a two dimensional operational definition of  consciousness for clinical purposes.  Assessments of  awareness use behavioral and neurophysiological (fMRI or EEG) protocols in order to gauge how patients perform on various cognitive functions.   Assessments of wakefulness are based on metabolic markers (if reporting is not possible) such as glucose uptake in the brain, captured using PET scans \cite{bodart2017measures}.   As such a clinically-oriented   definition of consciousness enables  classification of closely associated states  and  disorders of consciousness into clusters on a bivariate scale with awareness and wakefulness on orthogonal axes.  Under healthy conditions, these two levels are almost linearly correlated, as in conscious wakefulness (high arousal and high awareness) or in deep sleep (low arousal and low awareness). However,  in pathological states, wakefulness without awareness can be observed in the vegetative state \cite{laureys2004}, while transiently reduced awareness is  observed following seizures \cite{blumenfeld2012impaired}. Patients in the minimally conscious state show intermittent and limited non-reflexive and purposeful behavior \cite{giacino2002minimally}, \cite{giacino2004vegetative}, whereas patients with hemi-spatial neglect display reduced awareness of stimuli contralateral to the side where brain damage has occurred \cite{parton2004hemispatial}.   Given the aforementioned scales for labeling states and disorders of consciousness, the crucial question is how should one quantify awareness and wakefulness from neurophysiological data? This is particularly useful for non-communicative patients such as those in coma or states of minimal wakefulness. For this reason,  several dynamical complexity measures have been developed.  In this article, we first describe  theoretically-grounded complexity measures and the challenges one faces when applying these measures to realistic brain data.  We then  outline   alternative empirical approaches to classify states and disorders of consciousness. We end with a discussion on how these two approaches might inform each other. 

\section{Measures of Integrated Information } 
Dynamical complexity measures are designed to capture both, network topology as well as causal dynamics.  The most prominent among these is integrated information, denoted as $\Phi$.   This was first introduced in  \cite{tononi1994} and is defined as the quantity of information generated by a network as a whole,  over and above that of its parts, taking into account the system's  causal dynamical interactions.  This reflects the intuition going back to William James that conscious states are integrated, yet diverse.  $\Phi$  seeks to operationalize this intuition in terms of complexity, stating that complexity arises from simultaneous integration and differentiation of the network's structure as well as dynamics. Differentiation refers to functional specialization of neural  populations, while integration, as a complementary design principle, results in distributed coordination among neural populations.   This interplay generates integrated yet diversified information believed to  support  cognitive and behavioral states.  The earliest proposals defining integrated information were made in    \cite{tononi1994}, \cite{tononi2003} and \cite{tononi2004}.  Since then, considerable progress has been made towards the development of a normative theory as well as applications  of integrated information  \cite{bt}, \cite{bs}, \cite{tononi2012integrated},  \cite{arsiwalla2013iit}, \cite{oizumi2014},  \cite{arsiwalla2016computing}, \cite{arsiwalla2016high},  \cite{krohn2016computing},   \cite{tegmark2016improved}, \cite{arsiwalla2017spectral}.   
The core idea of integrated information as a whole versus parts quantity has been formalized in several distinct information measures  such as neural complexity \cite{tononi1994}, causal density \cite{seth2005causal}, $\Phi$ from integrated information theory: IIT 1.0, 2.0 \& 3.0 \cite{tononi2004}, \cite{bt}, \cite{oizumi2014}, stochastic interaction \cite{wennekers2005stochastic}, \cite{ay2015information},   stochastic integrated information \cite{bs}, \cite{arsiwalla2013iit}, \cite{xda2016global}  and synergistic $\Phi$ \cite{griffith2014quantifying}, \cite{griffith2014principled}.   \textbf{Table \ref{T1}} summarizes these measures along with corresponding information metrics upon which they have been based.  

\begin{table}[h!] 
\centering 
\caption{Theoretical complexity measures alongside their corresponding information metrics. }
\label{T1}
\begin{tabular}{|c|c|}
        \hline   
{\bf Integrated Information Measures}  & {\bf  Information Metrics}  \\  \hline  \hline 
Neural Complexity & Mutual Information (MI) \\  \hline
Causal Density & Granger Causality (GC) \\  \hline 
Stochastic Interaction & Kullback-Leibler Divergence (KLD)  \\  \hline 
IIT 1.0 \& 2.0 & KLD  \\  \hline  
Stochastic Integrated Information & MI or KLD  \\   \hline  
IIT 3.0 & Earth Mover's Distance \\      \hline   
Synergistic $\Phi$ & Synergistic Information \\  \hline  
\end{tabular}
\end{table}

However, computing integrated information for large neurophysiological datasets has been challenging due to both, computational difficulties and limits on domains where   these measures can be implemented.  For instance, many of these measures use the minimum information partition of the network. This involves evaluating a large number of network configurations (more precisely, the Bell number), which makes their computational cost extremely high for large networks.   As for domains of applicability, the measure of \cite{bt} has been formulated for discrete-state, deterministic, Markovian systems with the maximum entropy distribution. On the other hand, the measure of \cite{bs}  has been devised to continuous-state, stochastic, non-Markovian systems and in principle, admits dynamics with any empirical distribution (although in practice, it is easier to use assuming Gaussian distributions). The formulation in \cite{bs} is based on mutual information, whereas \cite{bt} uses a measure based on the Kullback-Leibler  divergence. Note however, that in some cases the measure of \cite{bs} can take negative values and that complicates its interpretation. The Kullback-Leibler based definition  computes the information generated during state transitions and  remains positive in the regime of stable dynamics. This gives it a natural interpretation as an integrated information measure. Both measures \cite{bt}, \cite{bs} make use of a normalization scheme in their formulations. Normalization inadvertently  introduces ambiguities in computations. The normalization is actually used for the purpose of determining the partition of the network that minimizes the integrated information, but a normalization dependent choice of partition ends up influencing the value and interpretation of  $\Phi$.  An alternate measure based on the Earth  Mover's distance was proposed in \cite{oizumi2014}. This does away with the normalization  problem (though the current version is not formulated for continuous-state variables). However,  the formulation of  \cite{oizumi2014}  lies outside the scope of standard information theory and  is still  difficult for performing computations on large networks. 

More recently, these issues have been addressed in  \cite{xda2016global}, using a formulation of stochastic integrated information based on the Kullback-Leibler divergence between the conditional multivariate distribution on the set of network states versus the corresponding factorized distribution over its parts, while implementing the maximum information partition instead of the minimum information partition.  Using this formulation, $\Phi$ can be computed for large-scale networks with linear stochastic  dynamics, for both, attractor as well as non-stationary states \cite{xda2016global}  (for network simulations  see  \cite{2791},  \cite{arsiwalla2015connectomics},   \cite{arsiwalla2015network}).  This work also demonstrated  the first  computation of  $\Phi$ for the resting-state human brain connectome.  The   connectome network is estimated  from cortical white matter tractography data,  comprising 998 voxels (nodes) with approximately 28,000 weighted symmetric connections  \cite{Hagmann2008}.  \cite{xda2016global} show that the dynamics and topology of the healthy resting-state brain generates greater information complexity than a (weight-preserving) random rewiring of the same network.  Even though this formulation of stochastic integrated information was successfully implemented for the human cerebral connectome, a network of  998 nodes and about 28,000 edges, it was limited to linearized dynamics. This is well-defined in the vicinity of attractor states such as the resting-state, however, it would be desirable to extend this formulation to include  non-linearities existing in brain dynamics.

\section{Empirical Measures   } 
Ideally, integrated information was intended as a measure of awareness, one that could  account   for informational differences between states and also  disorders of consciousness. However, as described above,  for realistic brain dynamics and physiological data that task has in fact proven difficult.  On the other hand,  the basic conceptualization of consciousness in terms of integration and differentiation of causal information has motivated several  empirical measures  that seek to classify consciousness-related disorders from  patient data.  For example,  \cite{barrett2012granger}  investigated changes in conscious levels using Granger Causality (GC) as a causal connectivity measure.  Given two stationary time-series signals, Granger Causality  measures  the extent to which the past of one  assists in predicting the future of the other, over and above the extent to which the past of the latter already predicts its own future  \cite{granger1969investigating}, thus quantifying causal  relations between  two signaling sources.   This was tested using   electroencephalographic (EEG)  data from subjects undergoing propofol-induced anesthesia, with signals source-localized to the anterior and posterior cingulate cortices. \cite{barrett2012granger}  found a significant increases in bidirectional GC in most subjects during loss of consciousness, especially in the beta and gamma frequency ranges.  Another useful measure of causal connectivity is transfer entropy, which extends Granger causality to the non-Gaussian case. However, so far this has only been implemented  on neuronal cultures by  \cite{wibral2014directed}  and holds future potential as a clinically relevant measure.  Yet another  measure that has already proven useful as a clinical classifier of conscious levels is the Perturbational Complexity Index (PCI), which was introduced by \cite{casali2013theoretically}  and tested on TMS-evoked potentials measured with EEG. PCI is calculated by perturbing the cortex with transcranial magnetic stimulation (TMS) in order to engage distributed interactions in the brain and then  compressing the resulting  spatiotemporal EEG  responses to measure their algorithmic complexity, based on the Lempel-Ziv compression. For a given segment of EEG  data, the Lempel-Ziv algorithm quantifies complexity by counting the number of distinct patterns   in the data. For example, this can be  proportional to the size of a computer file after applying a data compression algorithm. Computing the Lempel-Ziv compressibility requires binarizing the time-series data, based  either on event-related potentials or with respect to a given threshold.   Using PCI,  \cite{casali2013theoretically}  were able to discriminate levels of consciousness during wakefulness, sleep, and anesthesia, as well as in patients who had emerged from coma and recovered a minimal level of consciousness.  Later, the Lempel-Ziv complexity was also used by \cite{schartner2015complexity}  on spontaneous high-density EEG data recorded from subjects undergoing propofol-induced anesthesia. Once again, a robust decline in complexity was observed during anesthesia.  These are complexity measures based on data compression algorithms.  A qualitative comparison between a data compression measure  inspired by PCI and   $\Phi$ was made in  \cite{virmani2016compression}. While compression-based measures  do seem to capture certain aspects of  $\Phi$, the exact relationship between the two is not completely clear.  Nonetheless, these empirical measures have been  useful for clinical purposes,  in terms of broadly discriminating disorders of consciousness.  Another relevant complexity measure  is the weighted symbolic mutual information (wSMI),  introduced by  \cite{king2013information}.  This is a measure of global information sharing across brain areas. It evaluates the extent to which two EEG channels present nonrandom joint fluctuations, suggesting that they share common sources. This is done by first transforming continuous signals into discrete symbols, and subsequently  computing the joint probabilities of symbol pairs between two EEG  channels. Before computing the symbolic mutual information between two time-series signals, a weighting is introduced to disregard conjunctions of identical or opposite-sign symbols from the two signal trains as that could potentially arise from common-source artifacts.  In \cite{king2013information}  wSMI was estimated for   181 EEG recordings from awake but noncommunicating patients diagnosed in various disorders of consciousness (including 143 from patients in vegetative and minimally conscious states). This measure of information sharing was found to systematically increases with consciousness. In particular, it was able to distinguish patients in the   vegetative state, minimally conscious state, and fully conscious state.  
In \textbf{Table \ref{T2}} we summarize the above empirical measures along with their domains of application.  

\begin{table}[h!]   
\centering 
\caption{Empirical complexity measures alongside their tested domains of application. }
\label{T2}    
\begin{tabular}{|c|c|}   
        \hline   
{\bf Empirical Measures}  &  {\bf  Tested Application Domains}  \\ \hline  \hline 
Granger Causality  & Wakefulness vs propofol-induced   \\   
 &   anesthesia using EEG \\  \hline  
Perturbational Complexity Index & Wakefulness, sleep, anesthesia, coma \& minimal  \\
 &  consciousness using TMS-evoked EEG  \\  \hline  
Lempel-Ziv Complexity & Wakefulness vs propofol-induced   \\
 &  anesthesia using EEG  \\    \hline  
Weighted Symbolic Mutual Information &  Vegetative, minimally conscious \&  \\
 & fully conscious states using EEG \\     \hline 
\end{tabular}
\end{table}

\section{Discussion }
The paradigm-shifting proposal that consciousness might be measurable in terms of the  information generated by  causal dynamics of the brain as a whole, over the sum of its parts, has led to precise quantitative formulations of information-theoretic complexity measures.  These measures seek to operationalize the intuition that the  complexity associated to consciousness arises from simultaneous integration and differentiation of the brain's structural and dynamical hierarchies.  However,  progress in this direction has faced  practical challenges such as high computational cost upon scaling with network size. This is especially true with regard to realistic   neuroimaging or physiological datasets. Even in the approach of \cite{xda2016global}, where both, the scaling and normalization problem have been solved, the formulation is still applicable only to linear dynamical systems.  A possible way to extend this formulation to non-linear systems such as the brain might be to first solve the Fokker-Planck equations for these systems (as probability distributions will no longer remain Gaussian) and subsequently estimate entropies and conditional entropies numerically to compute $\Phi$.  Another solution to the problem might be to construct statistical estimators for the   covariance matrices from data and then compute $\Phi$.  

In the meanwhile, for clinical purposes, it has been useful to consider empirical  complexity  measures, which serve as classifiers that very broadly discriminate states of consciousness, such as between wakefulness and anesthesia or broadly between disorders of consciousness. However, these measures do not strictly correspond to integrated information. Some of them are based on signal compression, which does capture differentiation, though not directly integration.   So far these methods have been applied on the scale of EEG datasets. One has yet to demonstrate their computational feasibility for  larger datasets  (which might only be a matter of time though). All in all,  bottom-up approaches suggest important features that might help inform or constrain implementations of principle-based approaches. However, the latter are indispensable for ultimately understanding causal aspects of information generation and flow in the brain. 
  
This article is intended as a lookup table spanning the landscape of both,  theoretically-motivated as well as empirically-based complexity measures used in current consciousness research.  Even though, for the purpose of this article, we have treated complexity as a global correlate of consciousness, there are indications that multiple complexity types, based on cognitive and behavioral control, might be important for a more precise classification of various  states of consciousness  \cite{bayne2016there},  \cite{xda2017morpho}.  This latter observation alludes to the need for an integrative systems approach to   consciousness research, one that is grounded in cognitive architectures and helps understand control mechanisms underlying systems level neural information processing  \cite{conscious12016}.

\subsubsection*{Acknowledgments.}  This work has been supported by the European Research Council's CDAC project: "The Role of Consciousness in Adaptive Behavior: A Combined Empirical, Computational and Robot based Approach" (ERC-2013- ADG 341196).  

\bibliographystyle{splncs03}
\bibliography{test2}

\begin{thebibliography}{10}
\providecommand{\url}[1]{\texttt{#1}}
\providecommand{\urlprefix}{URL }

\bibitem{arsiwalla2013iit}
Arsiwalla, X.D., Verschure, P.F.M.J.: Integrated information for large complex
  networks. In: The 2013 International Joint Conference on Neural Networks
  (IJCNN). pp. 1--7 (Aug 2013)

\bibitem{2791}
Arsiwalla, X.D., Betella, A., Bueno, E.M., Omedas, P., Zucca, R., Verschure,
  P.F.: The dynamic connectome: A tool for large-scale 3d reconstruction of
  brain activity in real-time. In: ECMS. pp. 865--869 (2013)

\bibitem{arsiwalla2015connectomics}
Arsiwalla, X.D., Dalmazzo, D., Zucca, R., Betella, A., Brandi, S., Martinez,
  E., Omedas, P., Verschure, P.: Connectomics to semantomics: Addressing the
  brain's big data challenge. Procedia Computer Science  53,  48--55 (2015)

\bibitem{conscious12016}
Arsiwalla, X.D., Herreros, I., Moulin-Frier, C., Sanchez, M., Verschure, P.F.:
  Is Consciousness a Control Process?, pp. 233--238. IOS Press, Amsterdam
  (2016)

\bibitem{arsiwalla2017spectral}
Arsiwalla, X.D., Mediano, P.A., Verschure, P.F.: Spectral modes of network
  dynamics reveal increased informational complexity near criticality. Procedia
  Computer Science  108,  119--128 (2017)

\bibitem{xda2017morpho}
Arsiwalla, X.D., Moulin-Frier, C., Herreros, I., Sanchez-Fibla, M., Verschure,
  P.F.: The morphospace of consciousness. arXiv preprint arXiv:1705.11190
  (2017)

\bibitem{arsiwalla2016computing}
Arsiwalla, X.D., Verschure, P.: Computing Information Integration in Brain
  Networks, pp. 136--146. Springer International Publishing, Cham, Switzerland
  (2016)

\bibitem{arsiwalla2016high}
Arsiwalla, X.D., Verschure, P.F.M.J.: High Integrated Information in Complex
  Networks Near Criticality, pp. 184--191. Springer International Publishing,
  Cham, Switzerland (2016)

\bibitem{xda2016global}
Arsiwalla, X.D., Verschure, P.F.: The global dynamical complexity of the human
  brain network. Applied Network Science  1(1), ~16 (2016)

\bibitem{arsiwalla2015network}
Arsiwalla, X.D., Zucca, R., Betella, A., Martinez, E., Dalmazzo, D., Omedas,
  P., Deco, G., Verschure, P.: Network dynamics with brainx3: A large-scale
  simulation of the human brain network with real-time interaction. Frontiers
  in Neuroinformatics  9(2) (2015)

\bibitem{ay2015information}
Ay, N.: Information geometry on complexity and stochastic interaction. Entropy
  17(4),  2432--2458 (2015)

\bibitem{bt}
Balduzzi, D., Tononi, G.: Integrated information in discrete dynamical systems:
  motivation and theoretical framework. PLoS Comput Biol  4(6),  e1000091
  (2008)

\bibitem{barrett2012granger}
Barrett, A.B., Murphy, M., Bruno, M.A., Noirhomme, Q., Boly, M., Laureys, S.,
  Seth, A.K.: Granger causality analysis of steady-state
  electroencephalographic signals during propofol-induced anaesthesia. PloS one
   7(1),  e29072 (2012)

\bibitem{bs}
Barrett, A.B., Seth, A.K.: Practical measures of integrated information for
  time-series data. PLoS Comput Biol  7(1),  e1001052 (2011)

\bibitem{bayne2016there}
Bayne, T., Hohwy, J., Owen, A.M.: Are there levels of consciousness? Trends in
  cognitive sciences  20(6),  405--413 (2016)

\bibitem{blumenfeld2012impaired}
Blumenfeld, H.: Impaired consciousness in epilepsy. The Lancet Neurology
  11(9),  814--826 (2012)

\bibitem{bodart2017measures}
Bodart, O., Gosseries, O., Wannez, S., Thibaut, A., Annen, J., Boly, M.,
  Rosanova, M., Casali, A.G., Casarotto, S., Tononi, G., et~al.: Measures of
  metabolism and complexity in the brain of patients with disorders of
  consciousness. NeuroImage: Clinical  14,  354--362 (2017)

\bibitem{casali2013theoretically}
Casali, A.G., Gosseries, O., Rosanova, M., Boly, M., Sarasso, S., Casali, K.R.,
  Casarotto, S., Bruno, M.A., Laureys, S., Tononi, G., et~al.: A theoretically
  based index of consciousness independent of sensory processing and behavior.
  Science translational medicine  5(198),  198ra105--198ra105 (2013)

\bibitem{giacino2004vegetative}
Giacino, J.T.: The vegetative and minimally conscious states: consensus-based
  criteria for establishing diagnosis and prognosis. NeuroRehabilitation
  19(4),  293--298 (2004)

\bibitem{giacino2002minimally}
Giacino, J.T., Ashwal, S., Childs, N., Cranford, R., Jennett, B., Katz, D.I.,
  Kelly, J.P., Rosenberg, J.H., Whyte, J., Zafonte, R., et~al.: The minimally
  conscious state definition and diagnostic criteria. Neurology  58(3),
  349--353 (2002)

\bibitem{granger1969investigating}
Granger, C.W.: Investigating causal relations by econometric models and
  cross-spectral methods. Econometrica: Journal of the Econometric Society pp.
  424--438 (1969)

\bibitem{griffith2014principled}
Griffith, V.: A principled infotheoretic$\backslash$ phi-like measure. arXiv
  preprint arXiv:1401.0978  (2014)

\bibitem{griffith2014quantifying}
Griffith, V., Koch, C.: Quantifying Synergistic Mutual Information, pp.
  159--190. Springer Berlin Heidelberg, Berlin, Heidelberg (2014),
  \url{http://dx.doi.org/10.1007/978-3-642-53734-9\_6}

\bibitem{Hagmann2008}
Hagmann, P., Cammoun, L., Gigandet, X., Meuli, R., Honey, C.J., Wedeen, V.J.,
  Sporns, O.: {Mapping the Structural Core of Human Cerebral Cortex}. PLoS
  Biology  6(7), ~15 (2008)

\bibitem{king2013information}
King, J.R., Sitt, J.D., Faugeras, F., Rohaut, B., El~Karoui, I., Cohen, L.,
  Naccache, L., Dehaene, S.: Information sharing in the brain indexes
  consciousness in noncommunicative patients. Current Biology  23(19),
  1914--1919 (2013)

\bibitem{krohn2016computing}
Krohn, S., Ostwald, D.: Computing integrated information. arXiv preprint
  arXiv:1610.03627  (2016)

\bibitem{laureys2005}
Laureys, S.: The neural correlate of (un) awareness: lessons from the
  vegetative state. Trends in cognitive sciences  9(12),  556--559 (2005)

\bibitem{laureys2004}
Laureys, S., Owen, A.M., Schiff, N.D.: Brain function in coma, vegetative
  state, and related disorders. The Lancet Neurology  3(9),  537--546 (2004)

\bibitem{oizumi2014}
Oizumi, M., Albantakis, L., Tononi, G.: From the phenomenology to the
  mechanisms of consciousness: integrated information theory 3.0. PLoS Comput
  Biol  10(5),  e1003588 (2014)

\bibitem{parton2004hemispatial}
Parton, A., Malhotra, P., Husain, M.: Hemispatial neglect. Journal of
  Neurology, Neurosurgery \& Psychiatry  75(1),  13--21 (2004)

\bibitem{schartner2015complexity}
Schartner, M., Seth, A., Noirhomme, Q., Boly, M., Bruno, M.A., Laureys, S.,
  Barrett, A.: Complexity of multi-dimensional spontaneous eeg decreases during
  propofol induced general anaesthesia. PloS one  10(8),  e0133532 (2015)

\bibitem{seth2005causal}
Seth, A.K.: Causal connectivity of evolved neural networks during behavior.
  Network: Computation in Neural Systems  16(1),  35--54 (2005)

\bibitem{tegmark2016improved}
Tegmark, M.: Improved measures of integrated information. arXiv preprint
  arXiv:1601.02626  (2016)

\bibitem{tononi2004}
Tononi, G.: An information integration theory of consciousness. BMC
  neuroscience  5(1), ~42 (2004)

\bibitem{tononi2012integrated}
Tononi, G.: Integrated information theory of consciousness: an updated account.
  Arch Ital Biol  150(2-3),  56--90 (2012)

\bibitem{tononi2003}
Tononi, G., Sporns, O.: Measuring information integration. BMC neuroscience
  4(1), ~31 (2003)

\bibitem{tononi1994}
Tononi, G., Sporns, O., Edelman, G.M.: A measure for brain complexity: relating
  functional segregation and integration in the nervous system. Proceedings of
  the National Academy of Sciences  91(11),  5033--5037 (1994)

\bibitem{virmani2016compression}
Virmani, M., Nagaraj, N.: A compression-complexity measure of integrated
  information. arXiv preprint arXiv:1608.08450  (2016)

\bibitem{wennekers2005stochastic}
Wennekers, T., Ay, N.: Stochastic interaction in associative nets.
  Neurocomputing  65,  387--392 (2005)

\bibitem{wibral2014directed}
Wibral, M., Vicente, R., Lindner, M.: Transfer Entropy in Neuroscience, pp.
  3--36. Springer Berlin Heidelberg, Berlin, Heidelberg (2014)

\end{thebibliography}

\end{document}